\documentclass[aps,prl,twocolumn,superscriptaddress, fleqn, showpacs]{revtex4}

\usepackage{times,xspace}
\usepackage{amsbsy,amssymb,amsmath,bm}
\usepackage{graphicx,color,epsfig,rotate}

\def\bbbc{{\mathchoice {\setbox0=\hbox{$\displaystyle\rm C$}\hbox{\hbox
to0pt{\kern0.4\wd0\vrule height0.9\ht0\hss}\box0}}
{\setbox0=\hbox{$\textstyle\rm C$}\hbox{\hbox
to0pt{\kern0.4\wd0\vrule height0.9\ht0\hss}\box0}}
{\setbox0=\hbox{$\scriptstyle\rm C$}\hbox{\hbox
to0pt{\kern0.4\wd0\vrule height0.9\ht0\hss}\box0}}
{\setbox0=\hbox{$\scriptscriptstyle\rm C$}\hbox{\hbox
to0pt{\kern0.4\wd0\vrule height0.9\ht0\hss}\box0}}}}

\begin{document}

\title{Fractalisation drives crystalline states in a frustrated spin
system}

\author{Suchitra E. Sebastian}
\affiliation{Cavendish Laboratory, University of Cambridge, Madingley Road, Cambridge CB3 0HE, UK}
\author{N. Harrison}
\affiliation{NHMFL, MS-E536, Los Alamos National Laboratory, Los Alamos, New Mexico 87545, USA}
\author{P. Sengupta}
\affiliation{NHMFL, MS-E536, Los Alamos National Laboratory, Los Alamos, New Mexico 87545, USA}
\affiliation{Theoretical Division, Los Alamos National Laboratory, Los Alamos, New Mexico 87545, USA}
\author{C. D. Batista}
\affiliation{Theoretical Division, Los Alamos National Laboratory, Los Alamos, New Mexico 87545, USA}
\author{S. Francoual}
\affiliation{NHMFL, MS-E536, Los Alamos National Laboratory, Los Alamos, New Mexico 87545, USA}
\author{E. Palm}
\affiliation{National High Magnetic Field Laboratory, Tallahassee, Florida 32310, USA}
\author{T. Murphy}
\affiliation{National High Magnetic Field Laboratory, Tallahassee, Florida 32310, USA}
\author{N. Marcano}
\affiliation{Cavendish Laboratory, University of Cambridge,
Madingley Road, Cambridge CB3 0HE, UK}
\author{H. A. Dabkowska}
\affiliation{Department of Physics and Astronomy, McMaster University, Hamilton,
Ontario L8S 4M1, Canada}
\author{B. D. Gaulin}
\affiliation{Department of Physics and Astronomy, McMaster University, Hamilton,
Ontario L8S 4M1, Canada}
\affiliation{Canadian Institute for Advanced Research, Toronto, Ontario M5G 1Z8,
Canada}

\date{\today}
\begin{abstract}
We measure a sequence of quantum Hall-like plateaux at 1/$q$:
$9 \geq q \geq 2$ and $p$/$q$ = 2/9 fractions in the magnetisation
with increasing magnetic field in the geometrically frustrated spin system
SrCu$_{2}$(BO$_{3})_{2}$. We find that the entire observed sequence of
plateaux is reproduced by solving the Hofstadter problem on the system lattice
when short-range repulsive interactions are included, thus providing a sterling
demonstration of bosons confined by a magnetic and lattice potential mimicking
fermions in the extreme quantum limit.
\end{abstract}


\maketitle

Geometrical frustration in the spin dimer material SrCu$_{2}$(BO$_{3})_{2}$
\cite{Smith1, Kageyama1, Misguich1,Kodama1,Shastry1,Miyahara1,Kageyama2} leads
to a singlet Shastry-Sutherland groundstate at low magnetic fields,
but complex spin superstructures at higher fields. Our magnetisation measurements
reveal a fine substructure of quantum Hall-like plateaux at all $1/q$ ratios
$2<q<9$ and $p/q = 2/9$ in magnetic fields up to 85 T and temperatures down
to 29 mK, within the sequence of previously identified plateaux at
1/8, 1/4, and 1/3 of the saturated magnetization.
We identify this hierarchy of plateaux as a consequence of
confined bosons in SrCu$_{2}$(BO$_{3})_{2}$ mimicking the
high magnetic field fractalisation predicted by the Hofstadter
butterfly \cite{Hofstadter1} for fermionic systems.
Such an experimental realisation of the Hofstadter butterfly has not been
previously achieved in real interacting materials, given the unachievably
high magnetic flux densities or large lattice periods required.
By a theoretical treatment that includes short-range repulsion in the Hofstadter treatment,
stripe-like spin density-modulated phases are revealed in SrCu$_{2}$(BO$_{3})_{2}$
as emergent from a fluidic fractal spectrum.

The geometrically frustrated spin gap system SrCu$_{2}$(BO$_{3})_{2}$ is
unique in its orthogonal arrangement of spin dimers, resulting in an exact
direct singlet product incompressible groundstate, formally described by
Shastry and Sutherland \cite{Sutherland1}. Dimers comprising pairs of
$S$ = 1/2 spins on neighboring Cu$^{2+}$ ions bound by an intra-dimer
Heisenberg coupling, $J$, are coupled less strongly to orthogonal dimers
by an inter-dimer coupling $J$' (Fig. 1a inset), forming weakly coupled
layers within the tetragonal crystal structure. Magnetic susceptibility
and inelastic neutron scattering (INS) experiments measure a spin gap
of $\Delta $ = 34K, from which $J$/$J'\sim$1.47-1.67 ($J$ = 71.5-100K, $J $'= 43-68K)
\cite{Miyahara1,Nojiri1,Knetter1,Gaulin1} is estimated, placing the
system just within the predicted exact groundstate regime $J$/$J'\geq$1.35 \cite{Sutherland1,Koga1}.
In an applied magnetic field $H$, the groundstate incorporates a finite
density of spin triplets for which the groundstate configuration is no longer known.

We perform magnetisation measurements on oriented
SrCu$_{2}$(BO$_{3})_{2}$ crystals of $\sim $ 2mm x 2mm x 0.5mm using
cantilever magnetometry and $\sim $ 0.3mm x 0.3mm x 2mm using pulsed
fields, cut from larger single crystals grown using a self flux by
floating zone image furnace techniques \cite{Dabkowska1} and
characterised by INS \cite{Gaulin1} and Laue diffraction techniques.
Measurements up to 35T are made using a 10 $\mu$m thick CuBe torque
cantilever rotated in-situ to bring the crystalline $c$-axis within
2$^{\circ}$ of the applied magnetic field, and repeated on another sample up
to 45T using a 50 $\mu$m thick cantilever to confirm reproducibility and
facilitate comparison with previous magnetisation
measurements \cite{Kageyama1, Jorge1, Onizuka1, Kageyama3}. Torque
measurements were made in a portable dilution refrigerator in
continuous magnetic fields at NHMFL, Tallahassee - enabling a level
of sensitivity exceeding that possible in pulsed magnetic fields.
The measured torque is converted to absolute values of magnetisation
by multiplication by a constant rescaling factor and a small
quadratic background subtraction obtained on comparison with
measured values of pulsed field magnetisation \cite{Jorge1,
Onizuka1} (comparison shown in inset to Fig. 1a). In contrast to
ref. \cite{Levy1}, we find the magnetisation obtained from torque
measurements to correspond closely to that measured by
susceptometry in pulsed magnetic fields. Measurements between 38-85T
are made using a coaxially compensated wire-wound magnetometer
during the 10 ms insert magnet pulse in the NHMFL (Los Alamos)
multishot 100T magnet. While the short pulse duration necessarily
means a relatively large noise floor for these ultra high field
measurements, the identification of prominent features at these high fields
is facilitated, otherwise unachievable by other techniques.
Pulsed field measurements are made at $T$=500 mK during 2 magnet
shots on each of 2 different samples in order to confirm
reproducibility.

\begin{figure}[htbp]
\includegraphics[width=0.45\textwidth]{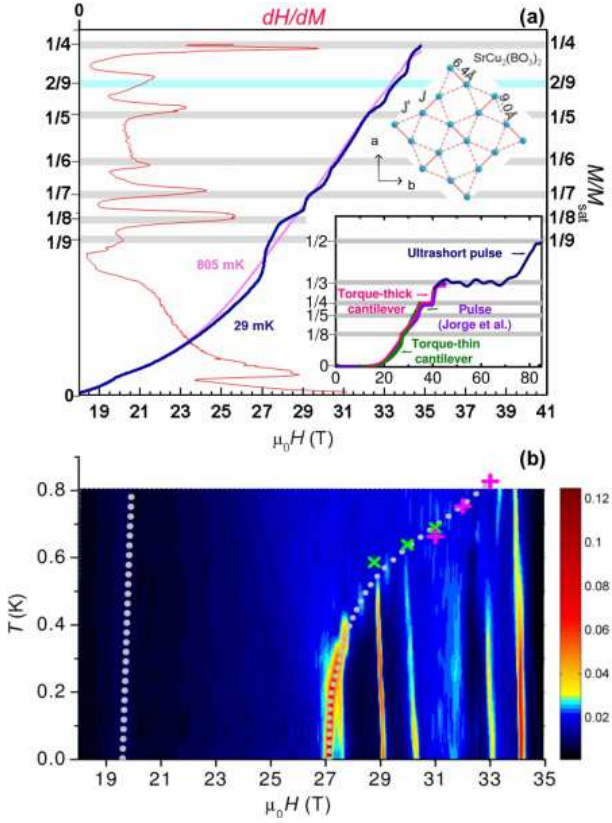}
\centering \caption {[Colour online] Experimentally measured magnetisation
of SrCu$_{2}$(BO$_{3})_{2}$ for $H \parallel c$ (a) Extracted from torque
measured using a 10$\mu$m cantilever in static magnetic fields up to 35 T
at representative temperatures. Plateaux at $m_{z}$/$m_{sat}=p$/$q$ are
indicated by peaks in inverse susceptibility d$H$/d$M$ (red). The lower
inset shows that low temperature data obtained using a 10$\mu$m cantilever up
to 35 T  (green), a 50$\mu$m cantilever up to 45 T (pink), ultrashort pulsed
field measurements from 38 to 85T (blue) and representative published pulsed
field magnetisation (violet) \cite{Jorge1} all overlay well. The upper inset shows the coupling of Cu$^{2+}$ spins within an SrCu$_{2}$(BO$_{3})_{2}$ layer (b) Contour plot of the temperature dependent differential susceptibility interpolated between
measurements using the most sensitive 10 $\mu$m cantilever at 29, 52, 160, 208,
294, 346, 450, 560, 628, 665, 705, 805 mK, with bright regions indicating
transitions between plateaux. Dotted lines represent the onset of a significant
triplet population, and the plateau phase boundary (guide to the eye). Crosses
represent heat capacity features (in pink) from Ref. \cite{Tsuji1}, and nmr features (in cyan) from Ref.~\cite{Takigawa1}.}
\label{fig1}
\end{figure}

On applying $H$ within 2$^{\circ}$ of the crystalline $c$-axis, a
discontinuous rise in magnetisation $m_{z}$ is observed above a
threshold magnetic field $\sim $ 19.5 T at 29 mK (Fig. 1a),
indicating the onset of significant triplet population
\cite{Kageyama1}. On further increasing $H$, a finely-spaced
sequence of plateau-like features (these are rounded due to
Dzyaloshinskii-Moriya terms \cite{Nojiri1,Cepas1} and thermal
smearing, we refer to them hereon simply as plateaux) appear in
$m_{z}$. Maxima in the inverse differential susceptibility
correspond to the midpoint of flat plateau regions, flanked by
thermally rounded transitions. Magnetisation plateaux values thus
located are found to occur in the sequence $m_{z}$/$m_{sat}$ = 1/9,
1/8, 1/7, 1/6, 1/5, 2/9 and 1/4 (within an error margin of 2$\%$),
where $m_{sat}$ is the saturation magnetisation (Fig. 1a). The
corresponding magnetic fields associated with each
plateau [midpoint (width)] are 27.4(0.4) T, 28.4(1.2) T, 29.7(0.8) T,
30.6(0.6) T, 32.6(0.8) T, 33.8(0.6) T, and 36(6) T respectively - the
boundaries being defined by maxima in the differential
susceptibility. The 1/3 plateau is located at 58(24) T from pulsed field
measurements (Fig. 1a inset). Fig. 1b shows the phase boundary corresponding
to melting temperatures of the cascade of ordered plateau
phases located from peaks in the differential
susceptibility. The increased sensitivity of the torque measurements
performed in continuous magnetic fields and significantly lower
temperatures enables the observance of additional plateaux previously
unobserved in magnetisation measurements at elevated temperatures
($T\geq$ 450 mK for $H \parallel$ c and $T\geq$ 80 mK for $H \parallel$ a)
using pulsed magnetic field measurements up to 70T
\cite{Kageyama1, Jorge1, Onizuka1, Kageyama3}, which reported only the
1/8, 1/4 and 1/3 plateaux. Subsequent measurements of nuclear magnetic resonance
(nmr) spectra~\cite{Takigawa1} performed down to temperatures of 0.19K find differences between
the spectral shape at field values 27.5 T, 28.7 T and 29.9 T, providing
corroboration for the fine plateaux
hierarchy at 1/9, 1/8 and 1/7 of the saturation magnetisation identified here by
torque measurements. While the $m_{z}$/$m_{sat
}$= 1/2 plateau was theoretically predicted in
Ref.\cite{Misguich1,Momoi1,Miyahara3}, we observe the first hint of its
existence in the pulsed field experiment reported here at fields
exceeding 80T. The 1/3 plateau is sufficiently stable and extended
in field range to enable its unambiguous identification even against
a relatively large noise floor (due to the short pulse duration in
the ultra high field 100T magnet), whereas experimental evidence for
the 1/2 plateau is more suggestive in nature.

We proceed to investigate the origin and character of the sequence of field-tuned
incompressible spin triplet configurations associated with the
observed magnetisation plateaux. The underlying confinement of spin-triplet
motion responsible for these features arises from the effect of geometrical
frustration in SrCu$_{2}$(BO$_{3})_{2}$. The minimal model for describing the
$S=1/2$ magnetic lattice is the Shastry-Sutherland Hamiltonian:
\begin{eqnarray}
\mathcal{H} &=& \sum_{ij}J_{ij}{\bf S}_i\cdot {\bf S}_j-g\mu_B\it{H}\sum_iS^z_i=
\sum_{ij}\frac{J_{ij}}{2}(b_i^\dag b_j+b_j^\dag b_i)
\nonumber \\
&+& J_{ij}(n_i-\frac{1}{2})(n_j-\frac{1}{2})-g\mu_B\it{H} \sum_i n_i
\label{Ham}
\end{eqnarray}
where $i$ and $j$ denote the lattice sites, ${\bf S}_i$ is the spin 1/2 operator
on site $i$, and $b^{\dagger}_i$, $b^{\;}_{i}$ are hard--core bosons
creation and annihilation operators at site $i$ that provide an alternative
description of a spin system via the Matsubara-Matsuda transformation \cite{Matsubara1}:
$S^{+}_i=b^{\dagger}_i$ and $S^{-}_i=b^{\;}_i$.
The off-site density-density interaction described by the last term
($n_{i }= b^{\dagger}_{i}b^{\;}_{i}$ is the number operator at site $i$)
corresponds to the Ising term, $S^z_i S^z_j$, in the spin representation.
We note that the local magnetization along the field direction
$\langle S^z_i \rangle=\langle n_i \rangle - 1/2$ corresponds to the particle density
in the bosonic representation.

The computation of $m_z(H)$ requires an approach that accurately
captures the underlying hierarchy of energy levels and allows to
obtain the groundstate energy for a quasi-continuum of densities -
unlike previously attempted $S^{z}$ = 1 triplet hardcore
bosonisation \cite{Momoi1, Miyahara3} and exact diagonalisation
techniques \cite{Kodama1, Miyahara2}. An alternate paradigm is
indicated by our experimental observation of plateaux at all $1/q$
ratios of $m_{sat}$ for 2$<q<$9 and $p$/$q$ = 2/9 -- reminiscent of
the quantum Hall effect \cite{MacDonald1, Avron1} described by
Landau level physics. Hence, we begin by adopting a fermionic
treatment in which the density-density interactions are assumed to
be irrelevant ($\langle n_i \rangle=\rho$ is uniform), as first
applied to SrCu$_{2}$(BO$_{3})_{2}$ by Misguich et al. in Ref.
\cite{Misguich1}. By using a Chern-Simons construction on the
lattice \cite{Fradkin1}, we can map the hard core bosons in
Eq.(\ref{Ham}) into spinless fermions: $b_j^\dag = f_j^\dag
e^{i\sum_{k\neq j}\mathrm{arg}(k,j)n_k}, b_j^\dag = e^{-i\sum_{k\neq
j}\mathrm{arg}(k,j)n_k}f_j $ where $arg(k,j)$ is the angle between
the relative vector, ${\bf r}_{k}-{\bf r}_j$, and an arbitrary direction. This transformation is equivalent to attaching a flux
quantum to each fermion such that the statistical phase generated
from fermionic pair-exchange is cancelled by the phase generated by
the flux quantum via the Aharonov-Bohm effect. After this
transformation, ${\mathcal{H}}$ becomes a model for gas of spinless
fermions moving on the same lattice and in the presence of a
non-local vector potential  $A_{ij}({\bf r}_i)$.  To simplify this
problem which is rendered intractable due to the  non--locality of
the vector potential, we make the approximation of a uniform flux
distribution generated by the statistical field
$H_s=4\Phi_0\frac{\rho}{a^2}$ \cite{Misguich1} ($\rho=\frac{1}{N_s}
\sum_{i} \langle n_i \rangle$ is the particle density and $N_s$ is
the number of sites). We thus realise a gas of \textit{interacting}
spinless fermions in a strong magnetic field $H_{s}$.

\begin{figure}[htbp]
\includegraphics[width=0.3\textwidth, angle=270]{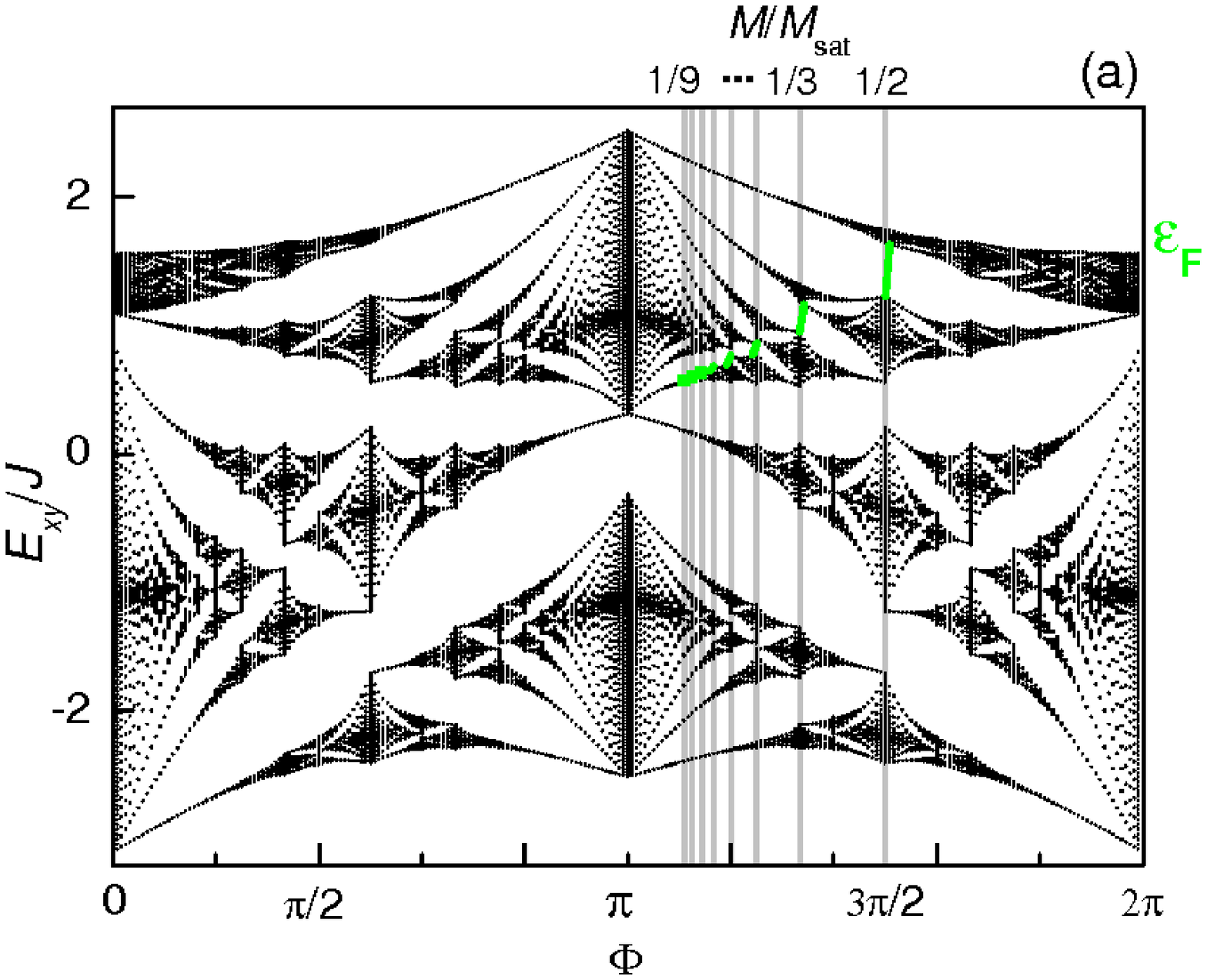}
\includegraphics[width=0.31\textwidth]{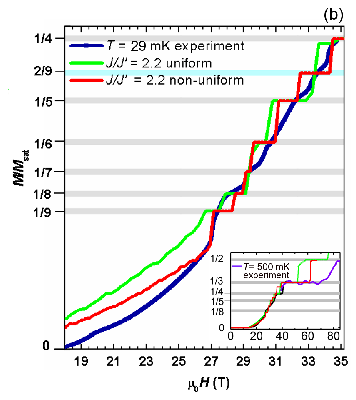}
\caption {[Colour online] Comparison between theoretical simulations and experiment.
(a) Computed fluid Hofstadter spectrum using a Chern-Simons treatment of the Shastry-Sutherland
lattice in the non-interacting limit for $J/J' $= 2.2 with the Fermi energy in green
- plateaux in the measured magnetisation correspond to gapped states. (a) Simulation
of $m_{z}$/$m_{sat}$ with $J$/$J'$ = 2.2 using system sizes $\sim$ 500 x 500: in
the uniform case neglecting interactions (green) and in the non-uniform case driven
by short-range repulsive interactions (red). $J/J'$ was selected to best match
the experimental data from a range of values incremented by $\pm$~0.1. Magnetisation
computed from the interaction-driven spectrum agrees well with experiment.}
\label{fig2}
\end{figure}

The motion of the bosons on the geometrically frustrated
Shastry-Sutherland lattice is therefore akin to orbital confinement
in two-dimensional quantum Hall systems by a (statistical) magnetic
field \cite{MacDonald1, Avron1}. Plateaux with a finite Hall
conductance arise when the chemical potential ($\mu=g_{c} \mu_B H$)
resides in a gap of the energy spectrum of fermions with average
density $\rho=m_{z}+1/2=a^{2} H_{s}/4\Phi _{o}$ ($a^{2}$ is the area
of the unit cell). The interplay of cyclotron (mean particle
separation `$\frac{a}{\rho^{1/2}}$') and lattice `$a$' length-scales
within this single particle band-filling picture was theoretically
captured by a Hofstadter butterfly of fractalised energy gaps in
1976 \cite{Hofstadter1}. Minimisation of the ground state energy
$E(m_{z})$ for each value of $H$ yields the dependence of $m_{z}$ on
$H$ in this non-interacting limit. Following this procedure for
SrCu$_{2}$(BO$_{3})_{2}$ (fractal spectrum shown in Fig. 2a),
theoretical $m_{z}$ curves are obtained (shown in Fig. 2b for $J$ =
70 K,$ J$/$J'=$2.2 optimised to match experimental plateaux
\cite{Exchange}), the shape of which agrees reasonably well with the
measured magnetisation and the values of observed plateaux (Fig.2b)
- an advance provided by this fermionic treatment over previous
models \cite{Kodama1, Miyahara1, Miyahara2, Momoi1, Miyahara3}.

\begin{figure}[h!tbp]
\includegraphics[width=0.48\textwidth]{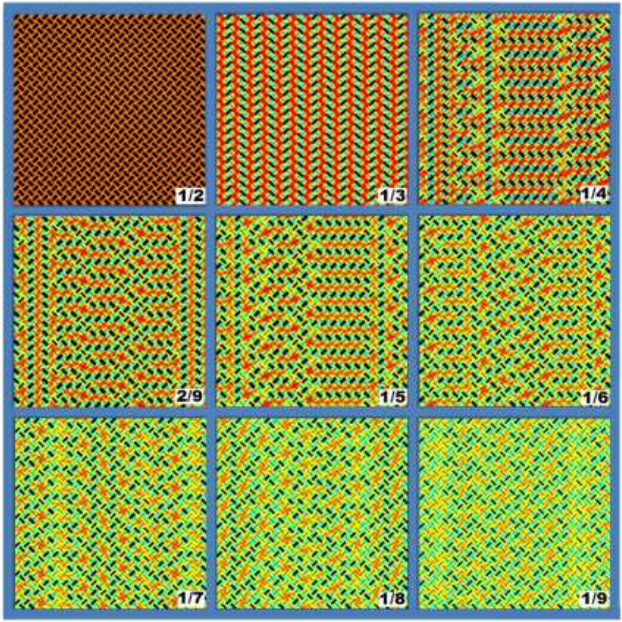}
\centering \caption {[Colour online] Minimal energy spin density profiles for each plateau. Calculated
using several different trial super unit cells (sizes $\sim $ (2-8) x
(2-36)) with different aspect ratios and on different finite lattices ($\sim
$ 500x500 original unit cells). Blue to red intensities represent
magnetisation from -1/2 to 1/2. Polarised dimers with chiefly singlet
contribution (negative magnetisation on one site) represented by thin black
lines, those with strong triplet contribution (positive magnetisation on
both sites) represented by thick black lines with widths proportional to
average magnetisation. A stripe super-structure modulates an approximately
uniform distribution of polarised-dimer clusters for $m_{z}$ /$m_{sat}\le $
1/3 plateaux.}
\label{fig3}
\end{figure}

The importance of density-density interactions, however, is apparent
from the experimental features unexplained by the uniform
Chern-Simons treatment. While this treatment captures most of the
measured plateau values and the overall shape of $m_{z}(H)$ by
neglecting density-density interactions, it fails to predict some
experimental plateaux. In addition, the uniform incompressible
liquid groundstate cannot explain the broken translational symmetry
observed by nmr measurements at the $m_{z}/m_{sat}$ = 1/8 plateaux
\cite{Kodama1} or the finite temperature  phase transitions measured
using heat capacity \cite{Tsuji1}. This suggests that off-site
repulsive interactions [see Eq.(\ref{Ham})] play an important role
in modifying groundstate character. To include this effect, we
consider an unconstrained and self-consistent mean field decoupling
$n_{i}n_{j}\simeq \langle n_{i}\rangle n_{j}+\langle n_{j}\rangle
n_{i}-\langle n_{i}\rangle \langle n_{j}\rangle$, thereby allowing
the mean value of the local density, $\langle n_{i}\rangle$, to
relax on each site of a super unit cell. Fig.\ref{fig3} shows the
spin density profile corresponding to the lowest energy solutions of
each of the $m_{z}$ /$m_{sat}=p/q$ plateaux, revealing that
stripe-like density modulations of the spin (except for the 1/2
plateau) mediated solely by short range interactions significantly
lower the energy.

While the predicted triplet superstructures in Fig.\ref{fig3} have some features in
common with spin configurations predicted in Refs.\cite{Kodama1, Momoi1, Miyahara3},
a qualitative difference is seen in the alternation of one-dimensional superstructures
with arrays of spiral-like clusters to accommodate coexisting microscopic length-scales
of the lattice `$a$' and the mean particle separation (analogous to the cyclotron radius)
`$\frac{a}{\rho^{1/2}}$'. This commensuration of lengthscales is characteristic to the
Hofstadter fractal spectrum, the gap hierarchy of which is inextricably linked to the
distinctive pattern of spin density modulation at each incompressible plateau state.
Starting with the fluidic spectrum (Fig. 2a), the gap structure is rearranged due to
a change in lattice potential as each $p$/$q$ incompressible plateau state stabilises
triplet stripes separated by an
average distance of $q$ lattice parameters (where $p$/$q$ is the density of $S^{z}$ = 1 triplets) \cite{Stripes}.
The role of interactions is further amplified by the quasi-degenerate energy level sub-spectrum at
lower $m_{z}$/$m_{sat}$ plateaux. For the associated low $p$/$q$ ratios with $q>>$1 (where $q$ determines the number of sub bands),
interactions stabilise a unique hierarchy of gaps from the fine gap substructure within a bounded spectrum (seen in Fig. 2a).

Spectacularly, the non-uniform solution corresponding to plateaux spin-density profiles in  Fig.\ref{fig3}
captures the entire sequence of measured magnetisation plateaux, including those observed at
$p/q$ = 1/7 and 2/9 which were not predicted by the uniform model -- comparison with
experiment is shown in Fig. 2b for a modified interaction-driven spectrum computed
for values of $J$ = 75.5 K and an optimised `effective' ratio $J$/$J'=$2.2 that best
fit experimental data. The remarkable correspondence of predicted and experimental
plateaux as well as the abrupt step in magnetisation preceding the onset of the
plateaux at $p/q$ = 1/9, and the background shape of the experimentally measured
magnetisation reveals a marked improvement over the uniform treatment.

We thus interpret the observed plateau states in
SrCu$_{2}$(BO$_{3})_{2}$ to represent the emergence of a new type of
stripe-like crystalline state resulting from the instability of the
incompressible Hall fluid to interactions. Unlike Wigner crystalline
states, the structure of these density-modulated states is seen to
be determined by the density-density susceptibility of the uniform
solution, as inferred from the $p$/$q$ fractions that are stabilised
as a function of field -- similar to the sequence which appears in
the fluid Hofstadter butterfly. We are now provided with a new
perspective from which to question the continued stability of
groundstates in metallic systems (composed of charged fermions) in
the formidably high equivalent experimental magnetic field limit.
The intermediate interaction regime in which the geometrically
frustrated SrCu$_{2}$(BO$_{3})_{2}$ system lies raises the
intriguing possibility of a dual groundstate in which crystalline
density-modulation retains topological properties of a Hall fluid,
enabling analogies to be made with `Hall crystals'
\cite{Tesanovich1} and field-induced charge-density wave states
\cite{Jain1}.


\begin{thebibliography}{1}
\bibitem{Exchange} This `effective' ratio $J/J'$ is higher than that
obtained from fits to low magnetic field susceptibility and INS data \cite{Miyahara1, Nojiri1,Knetter1, Gaulin1}
\end{thebibliography}
\end{document}